\documentclass{PoS}

\title{Monte-Carlo simulation of graphene in terms of occupation
numbers for the excitonic order parameter at hexagonal lattice.}

\ShortTitle{Graphene in terms of occupation numbers}

\author{\speaker{Oleg Pavlovsky}\thanks{Numerical calculations were performed at Supercomputing Center of
 Moscow State University}\\
        Institute for Theoretical Problems of Microphysics, Moscow State University and Institute for Theoretical and Experimental Physics, Moscow, Russia\\
        E-mail: \email{ovp@goa.bog.msu.ru}}

\author{Anna Sinelnikova\\
        Institute for Theoretical Problems of Microphysics, Moscow State University and Institute for Theoretical and Experimental Physics, Moscow, Russia\\
        E-mail: \email{sinel@goa.bog.msu.ru}}

\author{Maxim Ulybyshev\\
        Institute for Theoretical Problems of Microphysics, Moscow State University and Institute for Theoretical and Experimental Physics, Moscow, Russia\\
        E-mail: \email{ulybyshevl@goa.bog.msu.ru}}

\abstract{We present the results of the Monte-Carlo simulation of
graphene-like statistical model in terms of occupation numbers.
We study the problem of the phase transition in graphene to an
insulating phase. Only antiferromagnetic order parameter was
studied at the moment by means of Hybrid Monte-Carlo process on
the hexagonal lattice because of the sign problem for excitonic
order parameter in fermionic determinant. Therefore we have
studied the possibility of the phase transition to an excitonic
phase by means of simplified Monte-Carlo process on the hexagonal
lattice. We show that this phase, which corresponds to a charge
polarization of graphene sublattices, can appear at some values of
Coulomb electron-electron potentials. In this work we study the
dependence of the phase transition characteristics on the strength
of on-site electron-electron interaction. We show that excitonic
phase transition  is also impossible in suspended graphene.}

\FullConference{31st International Symposium on Lattice Field Theory - LATTICE 2013\\
        July 29 - August 3, 2013\\
        Mainz, Germany}

\begin{document}

\section*{Introduction}

%The unique mechanical and electronic properties of graphene
%attract a lot of attention from scientists and engineers at last
%years. Physics of graphene became an active
%area of research in the modern condensed matter physics.

One of the most essential problems in graphene physics is a
problem of electronic transport. Electronic excitations in
graphene strongly interact with each other. The strength of the
interaction is controlled by the substrate dielectric permittivity
so the strongest interaction is in suspended graphene. The
conductivity of suspended graphene has been an open problem for
many years. Many theoretical models \cite{Son:07:1, Khvech} have
predicted the existence of the mass gap in the case of free
graphene sheet which appears due to spontaneous breaking of
sublattice symmetry. This theoretical prediction have been
verified numerically within the framework of effective field
theory of graphene \cite{Lahde, Hands, Buividovich:2012uk}. Resent
experimental study \cite{Mayorov} of suspended graphene have shown
that this material remains in metallic state even without any
substrate. This discrepancy was solved in the paper \cite{arxiv}.
It was shown that if one takes into the account the screening of
the Coulomb potential at short distances, the tight-binding model of graphene
predicts the semimetal state for suspended graphene.

There are several order parameters and possible fermion
condensates which were discussed in the context of phase
transition phenomenon in graphene. For example, in the
papers\cite{Buividovich:2012uk,arxiv,Rebbi, Buividovich:2012nx}
the effect of spontaneous polarization of spins was studied. From
this point of view, the dielectric phase is an antiferromagnetic
state. The another possibility is excitonic condensation broadly
discussed both in theoretical papers and lattice simulations by
means of effective graphene field model. From microscopic point of
view,
 two sublattices of the hexagonal
lattice acquire opposite charges in this phase.
%\cite{Kittel}.

It is also important to emphasize that the study of this
excitonic order parameter within the framework of Hybrid Monte-Carlo
simulations (like it was done in \cite{arxiv} and \cite{
Buividovich:2012nx} for the antiferromagnetic order parameter) is
impossible now due to the sign problem in fermionic determinant. So in
this paper we have used simplified Monte Carlo method formulated in terms of
occupation numbers where this problem doesn't arise.

\section{Graphene model in terms of occupation numbers}

Let us consider an ideal monolayer graphene. Carbon atoms in it
form two dimensional hexagonal lattice. There is a 4-valency
carbon atom in every site of this lattice. Three electrons of each
carbon atom participate in chemical bonding ($\sigma$-bound) with
the neighboring atoms and the fourth electron ($\pi$-orbital)
causes the conductivity of graphene. We use the following
Hamiltonian to describe the electric properties of this material
\cite{Semenoff1}:
\begin{equation}
\label{fullH}  H = H_\kappa+H_H+H_C
=-\kappa\sum_{x,\rho_b,\sigma}\left(\psi^\dagger_{\sigma,x+\rho_b}
\psi_{\sigma, x} + h.c.\right) + { 1 \over 2} \sum_{x} q_x~V_{x,
x}~q_x + { 1 \over 2} \sum_{x\neq x'} q_x V_{x, x'} q_{x'} \, .
\end{equation}
Different terms of Hamiltonian correspond to different physical
phenomena.  $H_\kappa$ corresponds to hoppings from one site to
the nearest neighboring ones.
%\begin{equation}
%\label{hopping} H_\kappa=
%\kappa\sum_{x,\mu,\sigma}\left(\psi^\dagger_{\sigma,x+\mu}
%\psi_{\sigma, x} + \psi^\dagger_{\sigma, x} \psi_{\sigma,x+
%\mu}\right),
%\end{equation}
$\psi_{\sigma,x}^\dagger$ and $\psi_{\sigma,x}$ are the creation
and annihilation operators for the electrons at the site $x$ and
up or down spin, $\rho_b, \,\, b=1,2,3$ are the vectors between the site $x$ and
three its neighbors.  The term $H_H$ is the on-site Coulomb
interaction, $q_x$ is the operator of electrical charge for the site $x$:
 \begin{equation}
 q_x =
 1-\psi^\dagger_{\uparrow,x} \psi_{\uparrow,x} - \psi^\dagger_{\downarrow,x} \psi_{\downarrow,x}.
 \label{density}\end{equation}
``One'' in this formula represents the fact that the charge of the
lattice site is equal to 1 in the absence of electrons on $\pi$-orbitals.
%So we can transcribe on-site
%Coulomb interaction term $H_H$ using the charge operator $q_x$ as
%the following:
%\begin{equation}\label{hubbard}
%H_H= { 1 \over 2} \sum_{x}  q_x~V_{x, x}~q_x .
% \end{equation}
Finally, the last term in (1) is a Coulomb interaction of
electrons at different sites of the lattice.
%\begin{equation}\label{coulomb}
%H_C= { 1 \over 2} \sum_{x\neq x'} q_x V_{x, x'} q_{x'} = { 1 \over 2} \sum_{x\neq x'} q_x \frac{\alpha_{x,x'}}{|x-x'|} q_{x'} ,
%\end{equation}
$V_{x, x'}$ is a Coulomb potential screened by $\sigma$-orbitals at short distances. We
will discuss this phenomenon more properly in the next part of the
paper.

It's convenient to introduce electron and hole excitations.  Let
us consider the creation and annihilation operators for the
``electrons'' ($a$ operators) and ``holes'' ($b$ operators) at the
site $x$:
\begin{equation}
a^\dagger_x = \psi^\dagger_{x,\uparrow} \,\, , \,\,\,\,
a_x = \psi_{x,\uparrow} \,\, , \,\,\,\,
b^\dagger_x =\psi_{x,\downarrow} \,\,  , \,\,\,\,
b_x =\psi^\dagger_{x,\downarrow} \,\,  .
\label{ab}
\end{equation}
The charge operator in terms of operators $a_x$ and $b_x$ can be expressed as:
\begin{equation}
q_x=1-a^\dagger_x a_x - b_x\ b^\dagger_x =  b^\dagger_x b_x -
a^\dagger_x a_x.
\end{equation}
There are four different states for every site $x$:

1) The state
$\vert \cdot\,\cdot \rangle$. There is no electrons on the
$\pi$-orbital.

2) The states $\vert \uparrow\,\cdot \rangle$ and
$\vert \downarrow \cdot \rangle$. There is only one electron on
the $\pi$-orbital.

3) The state $\vert \uparrow\,\downarrow
\rangle$. There are two electrons on the $\pi$-orbital.

Each state in this set is also an eigenvector of ``electrons''
number operator and  ``holes'' number operator at the site $x$:
\begin{equation}
a^\dagger_x a_x \vert n_x, \, m_x \rangle =  m_x \vert n_x, \, m_x \rangle;
\,\,\,\,\,\,\,\,\,\,\,  b^\dagger_x b_x \vert  n_x, \, m_x \rangle = n_x \vert
 n_x, \, m_x \rangle \, ;\,\,\,\,\,\,\,\,\,\,\, n_x\,,\,m_x =0,1.\label{c}
\end{equation}

Let us choose ``vacuum'' state as $\vert \downarrow\,\cdot
\rangle$ with $n_x=0, \, m_x=0$ (there is no difference for Monte
Carlo method what state is called as ``vacuum''). Then $\vert
\uparrow\,\downarrow \rangle$ with  $n_x=1, \, m_x=0$ is
``electron'' excitation, $\vert \uparrow\,\cdot \rangle$ with
$n_x=1, \, m_x=1$ is ``electron+hole'' state, and
$\vert\cdot\,\cdot \rangle$ with $n_x=0, \, m_x=1$ is ``hole''
excitation. These vectors are eigenstates of charge operator with
eigenvalues $q_x=n_x-m_x$. As seen from the above, there are two
states with the same charges (equal to ``0''). One of them is the
``vacuum'' state $\vert \downarrow \, \cdot \rangle$, and the
other is the ``electron+hole'' state $\vert \uparrow \, \cdot
\rangle$. There is also one state with charge ``-1'' ($
\vert \downarrow \, \uparrow \rangle$) and one state with charge
``+1'' ($\vert \cdot \, \cdot \rangle$).

Now we can define the state of the whole system $ \vert S \rangle
$ as an antisymmetrized product of one-electron states: $ \vert S
\rangle=\vert \{n_x\}\{ m_x\}\rangle$, where $\{n_x\}$ and $\{
m_x\}$ are the distributions of occupation numbers for
''electrons'' and '' holes'' respectively at the whole lattice.

Let us discuss the meaning of three different parts of the full
Hamiltonian (\ref{fullH}). The values of electron-electron
interaction potentials for suspended graphene one can see in
\cite{Kac}. The energy scale in hopping part of the Hamiltonian is
defined by the value of $\kappa=2.7$ eV. The energy scale of
interaction part is defined by the Coulomb potential which is 9.3
eV in the case of on-site interaction. It is 3 times larger than
the hopping parameter $\kappa$. The eigenstates of particle number
operator are also eigenstates of Hubbard and Coulomb terms of
Hamiltonian (\ref{fullH}) but not of the $H_\kappa$ term. Because
of this fact and the fact that interaction  part of the
Hamiltonian gives possibly the main contribution to the energy, we
will neglect  $H_\kappa$. We can improve this method further by
perturbation theory if necessary.

So the partition function has the form:
\begin{eqnarray}
Z=\sum_{\{n\}\{m\}}e^{-\frac{1}{T}H(\{n\}\{m\})}
= \sum_{\{n\}\{m\}}\exp\left\{ -\frac{1}{2 T}\left( \sum_x V_{xx}(n_x-m_x)^2+
\sum_{x\neq y}V_{xy}(n_x-m_x)(n_y-m_y)\right)\right\} \nonumber
\end{eqnarray}
The average value of an observable quantity can be easily
calculated  if the eigenstates of particles number operators are
also eigenstates of the operator of this observable. In this case,
its average value can be represented as:
\begin{eqnarray}
\left< O \right> = \frac{ \sum_{\{q_x\}} O(\{q_x\}) e^{-\beta H}
\prod_x (1+\delta_{q_x,0})}{ \sum_{\{q_x\}}  e^{-\beta H}} ,\nonumber \\
 H = {1 \over 2 } \left( \sum_x V_{xx} (q_x)^2 + \sum_{x\neq y} V_{xy}(q_x)(q_y) \right),
\label{aver}
\end{eqnarray}
where $\beta = 1/T$, and $(1+\delta_{q_x,0})$ corresponds to the
existence of two different states with $q_x=0$.

\section{Simple model: we take into account only interactions in the 1st coordination radius}

Let us consider the model with the following partition function
\begin{equation}
Z =  \sum_{\{q_x\}} \prod_x (1+\delta_{q_x,0}) \exp^{-\beta H(\{q_x\})}, \, \, \, \,
\mbox{where} \, \, \, \,  H(\{q_x\}) =  {1 \over 2} \left( V_{00}
\sum_{x} q^2_x + V_{01} \sum_{x, \rho_b} q_x q_{x+\rho_b} \right).
\label{simp_ham}
\end{equation}
Only interaction with neighbouring sites is included into the
Hamiltonian.

The first term in Hamiltonian corresponds to the interaction
between electrons at one lattice site. The second term corresponds
to the Coulomb interaction between the nearest neighbours. The
vectors $\rho_b$, $b=1,2,3$ connect the nearest sites with each
other.

\begin{figure}
\centering \includegraphics[width=.3\textwidth]{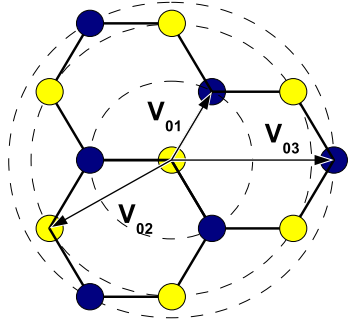}
\caption{The system with charge separation.}
\label{fig1}
\end{figure}

The relation between $V_{00}$ and $V_{01}$ is the key quantity
which defines the behaviour of the model. Let us consider the
system at large $\beta$, when the ground state plays the main role
in the partition function. We will consider firstly the limit
$V_{00} \gg V_{01}$. In this case the self energy contribution,
which corresponds to the on-site interaction term, dominates the
interaction between neighbouring sites. The configuration with
zero charges $q_x=0$ at all sites is the ground state    of the
system. In the opposite case $V_{00} \ll V_{01}$ the second term
in the Hamiltonian  (\ref{simp_ham})will dominate the first term
and the ground state is the configuration where sites at different
sublattices acquire opposite charges: $q_x = - q_{x+\rho_b}$. This
situation is illustrated at the figure \ref{fig1}, where different
charges are marked with different colors. Every term  $q_x
q_{x+\rho_b}$ gives a negative contribution into the energy. This
contributions leads to the negative energy of such ''chiral
domain'' in the limit $V_{01} \gg V_{00}$ and this configuration
becomes a ground state because the zero charges configuration has
zero energy. So, there is a critical value of $V^c_{00}$ where one
ground state is replaced by another one.

The value of  $V^c_{00}$ at large $\beta$ can be found from the
following considerations. Zero energy of the ''chiral domain''
configuration marks the point of the phase transition. We can calculate
the ''chiral domain'' energy per one lattice cell. One graphene
lattice cell contains two sites and tree links between sites. The
condition of zero energy per lattice cell can be written as
follows: $V_{00} = 3 \cdot V_{01}$.

It is impossible to find the critical $V^c_{00}$ analytically if
we take into account finite temperature and interaction at all
radii. So, we'll use Monte-Carlo to study this problem further in
the next section.

\section{Monte Carlo investigation of the phase diagram}

Let us turn to the numerical study of the models   (\ref{aver})
and (\ref{simp_ham}) by means of Monte-Carlo method. We have used
the heat bath algorithm for our calculations, taking into account
double weight of the states with zero charge $q_x=0$. We used
hexagonal lattice with $36 \times 36$ sites and periodical
boundary conditions.

We study the following observable to mark the phase transition:
\begin{equation}
 \langle O \rangle = \langle q_x q_{x+\rho_b} \rangle.
\end{equation}
This average is very convenient for the detection of the
transition between two vacua. Indeed, this quantity is close to
zero in case  disordered system. Otherwise, it is equal to $-1$ in
the chiral domain phase, because $q_x$ and $q_{x+\rho_b}$ belong
to different sublattices and have opposite charges.

We generated 100 statically independent configurations for every
value of  $V_{00}$ and $\beta=1/T$. We changed the value of
$V_{00}$ and hence the ratio between  $V_{01}$ and ${V_{00}}$. The
values of $V_{x y}$ at large distances correspond to ordinary
Coulomb interaction. Its strength is defined by the $V_{01}$
potential. So we vary the value of ${V_{00}}$ potential leaving
the other constant.

\begin{figure}
\hspace*{-4.0cm} \centering
\includegraphics[width=0.6\textwidth,angle=-90]{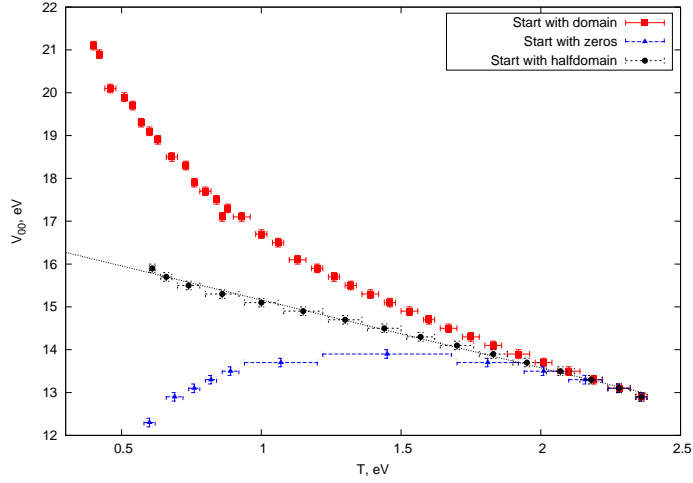}
\vspace{-70pt} \caption{Phase  diagram in the plane $(T, V_{00})$
for the model which takes into account only interaction  of the
nearest neighbours. Three different kinds of initial
configurations for termalization were used in the simulations.}
\label{fig2}
\end{figure}

At the beginning, we consider the simple model (\ref{simp_ham})
where only the nearest neighbours interact with each other. The
phase transition is marked by the maximum of derivation ${\partial
\langle O \rangle} \over { \partial T} $. The phase diagram is
shown in the figure \ref{fig2}. There are tree different curves
that corresponds to three different starts of Markov process
during Monte-Carlo calculation. The bottom curve corresponds to
start from configuration filled with zero charges. The top curve
corresponds to chiral domain starting configuration. The curve
between them corresponds to start from a special type of
configuration: half of the lattice is filled with chiral domain
and the rest of the lattice is filled with zero charges. The
results of Monte Carlo algorithm shouldn't depend on the initial
configuration for infinite thermalization time (infinite length of
the Markov process). But in practice it's impossible to provide
infinite time for thermalization. That is why we use that peculiar
configuration for start. In this case, the long termalization is
not necessary. We can only measure the change of the domain volume
to determine the phase which will be a result of the
thermalization process.

The accuracy of this method can be confirmed by the correspondence
between low-temperature extrapolation presented  in the the figure
\ref{fig2} and the theoretical value of the critical on-site
interaction potential $V^c_{00} = 3 \cdot V_{01}=16.5$~eV.

\begin{figure}
\hspace*{-4.0 cm}\centering
\includegraphics[width=0.6\textwidth,angle=-90]{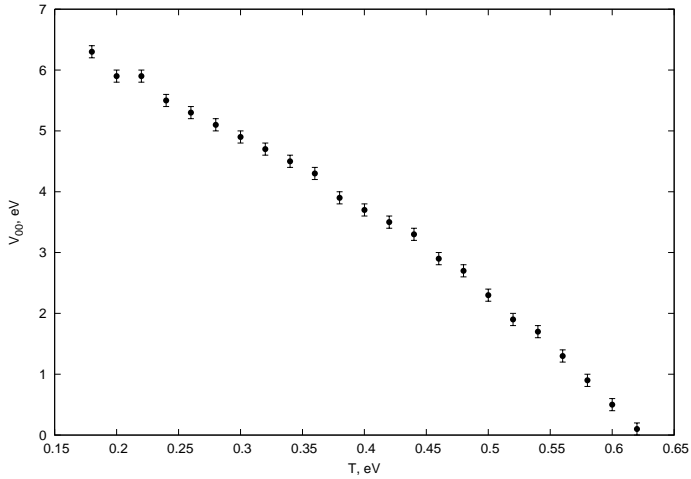}
\vspace{-70pt} \caption{Phase diagram in the plane $(T, V_{00})$
for the model with long range interaction. Start from halfdomain
configuration.} \label{fig3}
\end{figure}

Now let us investigate the system with interaction at all
coordination radii. In this case we again use the start from
lattice that is half-filled with domain. We detect the change of
the chiral domain volume. If it becomes smaller, the domain melt,
so the system turns into plasma. If domain grows, the systems
turns into the phase with ordered charges. The result one can see
in figure \ref{fig3}. The comparison of figures \ref{fig2} and
\ref{fig3} allows to trace the influence of Coulomb interaction at
large coordination radii.

\section*{Discussions and conclusion}

We investigated  the phase diagram in the plane $ (T , V_ { 00 })
$ for the excitonic phase transition at hexagonal  lattice. The
main observation is that the value of the on-site interaction
$V_{00}$ strongly influences the ground state of the system.
Decreasing of $V_{00}$ can lead to the change of  the system's
ground state from zero charges configuration to the configuration
with spontaneous charge separation between sublattices. The last
phase is a dielectric with nonzero excitonic condensate. In the
real graphene the $\sigma$ - orbital screening decreases the
on-site potential $V_{00}$. But this screening is not enough
strong for the phase transition even at zero temperature. The
phase diagram strongly depends also on the interaction at large
distances. So the real physical situation is a combination of two
effects: the on-site interaction screened by $\sigma$ - orbitals
and long range Coulomb interaction. The phase transition is very
sensitive to both these factors. We obtained that free graphene is
a conductor even at zero temperature. In order to achieve the
dielectric state with nonzero excitonic condensate, the $\sigma$ -
orbital screening should be stronger or something should block the
Coulomb interaction between electrons at large distances.

\end{document}